\documentclass[a4paper, 10pt, conference]{ieeeconf}      
\usepackage{graphicx}

\IEEEoverridecommandlockouts                              
\title{\LARGE \bf
Multi-Modal Continuous Valence And Arousal Prediction in the Wild Using Deep 3D Features and Sequence Modeling}

\author{\parbox{16cm}{\centering
    {\large Sowmya Rasipuram, Junaid Hamid Bhat and Anutosh Maitra}\\
    {\normalsize
    Accenture Technology Labs, Bangalore, India.\\
    }}
    \thanks{This work was not supported by Accenture Technology Labs, India. }
}

\begin{document}

\author{Sowmya Rasipuram, Junaid  Hamid Bhat and Anutosh Maitra\\ Accenture Technology Labs, Bangalore, India \\}
\pagestyle{plain}

\maketitle

\begin{abstract}
Continuous affect prediction in the wild is a very interesting problem and is challenging as continuous prediction involves heavy computation. This paper presents the methodologies and techniques used 
in our contribution to predict continuous emotion dimensions i.e., valence and arousal in ABAW competition on Aff-Wild2 database. Aff-Wild2 database consists of videos in the wild labelled for valence and arousal at frame level. 
Our proposed methodology uses fusion of both audio and video features (multi-modal) extracted using state-of-the-art methods.
These audio-video features are used to train a sequence-to-sequence model that is based on Gated Recurrent Units (GRU). We show promising results on validation data with simple architecture. The overall valence and arousal of the proposed approach is 0.22 and 0.34, which is better than the competition baseline of 0.14 and 0.24 respectively. 

\end{abstract}

\section{INTRODUCTION}
Affective Behavior Analysis in the Wild (ABAW) challenge consists of three tracks for predicting valence and arousal emotions, seven discrete emotions and Action Unit (AU) recognition. In this paper, we summarize our approach and results for predicting valence and arousal levels on Aff-Wild2 database. The database is released as a part of the challenge and annotations are provided at frame level for both valence and arousal levels \cite{kollias}. 

Aff-Wild2 consists of 545 videos with 2, 786, 201 frames. Some of these videos have left and right subjects and have been annotated for both separately. This is the largest audio-visual database available with intense annotations. It is an extension to the Aff-Wild database \cite{affwild}, \cite{cvpr}, \cite{kollias2017}. Kollias et al. reported concordance correlation coefficient (CCC) of 0.53 for valence and 0.44 for arousal using visual features \cite{kollias2018}.  In \cite{kollias2018_1}, the authors use adversarial networks for predicting valence and arousal values. In \cite{bmvc}, the authors use a multimodal approach based on CNN, RNN and GRU. In our work, we compare results that are presented in \cite{kollias}.

The training, validation and testing data splits have been given by the challenge separately for all tracks. In our approach, the main steps that we followed include a) Pre-processing for face detection to extract cropped and aligned face images from all videos, b) Extraction of audio features, c) Extraction of visual features, d) Training using deep learning methods. 

\subsection{Pre-processing}
The cropped face images for every video are provided as a part of the challenge. The number of frames in a video is determined using ffmpeg. The number of frames determined using ffmpeg did not match with the cropped face images provided. And hence, we used a different face detector to extract aligned faces. We used Openface, an opensource facial behavior analysis toolkit \cite{openface}. The number of face images obtained using Openface and ffmpeg matched exactly. It detects faces in robust environments when images are non-frontal, occluded and in low illumination conditions. It outputs faces of fixed dimension 112*112. These faces are used for further feature extraction. 

Audio features are extracted from the wav signal. The wav signal is extracted using ffmoeg, an opensource tool for audio processing. The wav files thus generated are usef for further audio feature extraction.

\subsection{Feature Extraction}
\subsubsection{Audio}
We extracted Mel-Frequency Cepstral Coefficients (MFCC) and Melspectrogram coefficients. Audio features are extracted using librosa\footnote{https://librosa.github.io/librosa/feature.html}.  These features were found to be effective for emotion recognition tasks \cite{yang}, \cite{badshah}. First, audio signal is split into N overlapping segments where N corresponds to the number of frames in video. The overlap is kept equal to one-half split. MFCC's form a cepstral representation where the frequency bands are not linear but distributed according to the mel-scale. The mfcc feature dimension obtained is 40. Melspectrogram (mel-frequency spectrogram) gives signal strength at various frequencies. We obtain a feature of dimension 128. The audio features are concatenated to form a feature vector of dimension 168. Frame-level audio features are then combined to form sequences of length 15 with an overlap of 5 frames for model training. 

 \begin{figure*}[ht]
    \centering
\includegraphics[width=6in,height=3in]{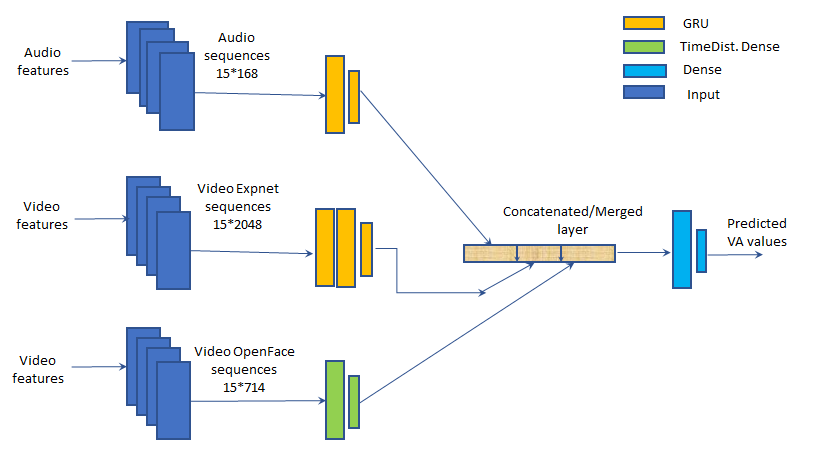}  
\caption{Block diagram of the proposed multi-modal model for Valence Arousal (VA) prediction}
    \label{fig:approach}
\end{figure*}

\subsubsection{Video}
Visual cues play an important role in recognizing human emotions. In this work, we automatically get a visual representation for each aligned face image using state of the art methods. We use ExpNet (Expression Net) proposed by Chang et al. \cite{Chang} to get a feature representation. ExpNet uses a landmark free approach to generate 3D expression coefficients based on CNN (Convolutional Neural Network). It is a deep learning based method that produces a 3D morphable model (3DMM) of the input face. This method not only captures emotion, but also gives position and shape. This approach is computationally efficient as it does not use any landmark detector that is redundant for this task. The authors show promising results for emotion recognition on benchmark datasets. Different features that can be extracted are:
\begin{enumerate}
     \item ExpNet - to estimate the emotion.
    \item Shape Net - estimate the 3D shape of the face.
    \item Face Pose Net - estimate the position of the face. 
\end{enumerate}

We are using only the emotion feature extractor to get a feature representation for every aligned face image. The aligned face images extracted using Openface are passed to ExpNet block to generate a feature vector of dimension $2048$. We extract features for all aligned face images and then combine frame-level representation to form a sequence level representation of dimension $15*2048$ with an overlap of $5$ frames. The sequence level features are used for model training. Along with using deep 3D ExpNet features, we also use features such as head pose, eye gaze, action unit intensities obtained from Openface. Openface generates a vector of dimension $714$ with the above features. We call these features as FacePose features. These frame-level features are also combined to form sequence level features of dimension $15*714$ with an overlap of 5 frames. 



\subsection{Experimental Details}
The three sets of features i.e., audio features, video features from ExpNet and FacePose features are processed separately before fusion. Direct fusion will lead to a computationally expensive model with huge number of parameters. And hence, we process them separately before fusion. Towards this fusion, we did several experiments on the hyperparameters. As mentioned previously, the sequence level features obtained by combining frame-level features are used in this step. The dimension of the sequence-level audio features is 15*168, sequence-level ExpNet features is 15*2048 and sequence-level FacePose features is 15*714. We implemented our codes in Python 3.7 and use Keras for deep learning framework. We run our model training on 8 core GPU. 


The audio sequences of dimension 15*168 are passed through two layers of GRU with 128 and 64 units respectively. We used a PReLU (Parametric ReLU) activation function with a dropout of 0.25 for each layer output. Video sequences from expnet of dimension 15*2048 are passed through three layers of GRU with 256, 256 and 64 units respectively. Each of the GRU layer is followed by an activation layer with PReLU activation function. Video sequences from OpenFace of dimension 15*714 are passed through two TimeDistributed Dense layers with dimension 128 and 64 respectively with a dropout of 0.25. All the layer outputs are followed by batch normalization.  

The 64 dimensional sequence output from the three blocks are concatenated and passed through two dense layers with 64 and 2 units respectively. The last dense layer is followed by tanh activation function as the valence/arousal values are in the range of [-1, +1]. We used mean squared loss function with rmsporp optimizer. The learning rate has been kept at 0.0001 after performing few experiments. Figure \ref{fig:approach} represents the overall approach in a snapshot. The same architecture has been tested by replacing GRU with Bidirectional LSTM's. The BiLSTM architecture gives similar results and are not presented here for repetition. As we use state-of-the-art methods for feature extraction that were based on deep networks and heavy training, we get promising results with our architecture. The results are validated on the validation data split provided to us. 

\begin{table}[]
    \centering
       \caption{Results on validation data}
    \label{tab:results}
    \begin{tabular}{c|c|c|c}
    \hline
        Features & Model & Valence & Arousal \\ 
        \hline
        Baseline & White paper & 0.14 & 0.24 \\
        Audio only & GRU layers & 0.14 & 0.21 \\
        video only (ExpNet) & GRU layers & 0.18 & 0.29 \\
        Audio+Video & Figure \ref{fig:approach} & \textbf{0.22} & \textbf{0.34} \\
        \hline
    \end{tabular}
 
\end{table}

Table \ref{tab:results} shows experimental results for predicting valence and arousal values using audio alone, video alone and audio-visual fusion. For audio-alone framework, the audio sequences are passed through two layers of GRU and two dense layers. Our best results (audio-only) show a concordance correlation coefficient (CCC) value of 0.14 for valence and 0.21 for arousal. When video features alone from expression net are considered, we observed best results with three layers of GRU followed by two dense layers.
we observed an improvement over the baseline. CCC value of 0.18 for valence and 0.29 for arousal is observed. When audio and video features are passed through the network shown in Figure \ref{fig:approach}, we observed the best CCC of 0.22 for valence and 0.34 for arousal. This indicates that fusion of features play a significant role to identify the emotion. The test results have been shared to the team for evaluation.

\section{Conclusion}
In this paper, we performed automatic multi-modal prediction of valence and arousal values on Aff-Wild2 database. One application of this framework can be used to identify the stress levels of people at workplace and provide suggestions for better health. Our automatic framework consisted of extraction of both audio and visual features. The visual features extracted using expression net and Openface are very suitable for predicting emotion. The multi-level architecture shows significant improvement of CCC using multi-modal cues. This work can further be extended by including other available datasets in training and more rigorous evaluation.

\section{Acknowledgements}

The authors gratefully acknowledge the contribution of reviewers' comments.


\end{document}